\documentclass{elsart}
\usepackage[]{latexsym,amsmath,amssymb}
\usepackage[]{epsfig}

\begin{document}

\begin{frontmatter}

\title{``Kerrr'' black hole:
the Lord of the String}

�\author{Anais Smailagic}
 \thanks[smaila]{email: anais@ts.infn.it}
 \address{Istituto Nazionale di Fisica Nucleare, Sezione di Trieste,
 Trieste, Italy }

 \author{Euro Spallucci}
  \thanks[euro]{email: spallucci@trieste.infn.it}
�  \address{Dipartimento di Fisica  dell' Universit\`a  di
Trieste, and Istituto Nazionale di Fisica Nucleare, Sezione di Trieste,
Trieste, Italy}

\begin{abstract}
Kerrr in the title is not a typo. The third ``r'' stands for \textit{regular}, in the 
sense of \textit{pathology-free} rotating black hole. We exhibit a long search-for, exact, Kerr-like, solution of the Einstein equations with novel features: i) no curvature ring singularity; ii) no ``anti-gravity'' universe with causality violating timelike closed
world-lines  ; iii) no ``super-luminal'' matter disk. \\
The ring singularity is replaced by a classical, circular, rotating string 
with Planck tension representing the inner engine driving the  rotation of all the surrounding
matter.\\
 The resulting geometry is \textit{regular} and \textit{smoothly} interpolates among inner Minkowski
space, borderline deSitter and outer Kerr universe. The key ingredient to
cure all unphysical features of the ordinary Kerr black hole is
the choice of a ``noncommutative geometry inspired'' matter source as the 
input for the Einstein equations, in analogy with spherically symmetric black holes
described in earlier works. 
  \end{abstract}
\end{frontmatter}

\section{Introduction}

Among the several black hole solutions of the Einstein equations, the Kerr geometry is without any doubts the most appropriate to fit the observational data showing that collapsed objects exhibit high angular momenta. Therefore, the complete and through understanding of its properties is crucial
for correct description of astrophysical objects. Furthermore, recent expectations are related to the possible outcome of LHC experiments, including production of micro black holes. \\
On the other hand, the history of  Kerr solution
is studded with technical difficulties in solving Einstein equations, accompanied with a complete ignorance of the appropriate matter source. The textbook procedure is based on the so-called ``vacuum solution'' method consisting in assuming  an ``ad-hoc`` symmetry for the metric and solving field equations with no source on the r.h.s. Integration constants are then determined comparing the weak-field limit of the solution with known Newtonian-like forms. While mathematically
correct this approach is physically unsatisfactory especially in General Relativity, where
basic postulate is that geometry is determined by the mass-energy distribution. Furthermore,
insisting on the vacuum nature of the solution leads to 
the presence of curvature singularities where the whole classical theory, i.e. General
Relativity, fails. In the Kerr geometry, there are further complications such as: 
 an anti-gravity region and causality violating closed time-like curves.
 These pathologies should not be present in a physically meaningful gravitational field. 
 A simple way out, is to replace the pathological vacuum region with a regular matter source.
 In a series of recent papers we have presented a spherically symmetric, regular, matter distribution  leading to both neutral and charge black hole solutions  with no
 curvature singularity 
 \cite{Nicolini:2005zi,Nicolini06,Ansoldi07,Spallucci:2008ez,Nicolini:2008aj,Casadio:2008qy},
 \cite{Nicolini:2009gw,Bleicher:2010qr,Rizzo:2006zb,Gingrich:2010ed}
 and traversable wormholes \cite{Garattini:2008xz}. Global
 structure and inner horizon stability for such a kind of geometries are currently under investigation
 \cite{Batic:2010vm,Arraut:2010qx}.
  The regularity of the metric follows from the presence of a
 \textit{minimal length} providing a universal cut-off for short-distance physics.  The idea that there should be a minimal distance is supported by many results in different approaches to quantum
 gravity \cite{sny,dw,Yoneya:1976pb}, \cite{Padmanabhan:1986ny,Padmanabhan:1987au,Padmanabhan:1988se},
 \cite{Ashtekar:1992tm,Rovelli:1992vv},
 \cite{Garay:1996sf,Garay:1994en,Fontanini:2005ik,Nicolini:2009dr,Modesto:2009qc}, 
 \cite{Rinaldi:2009ba,Rinaldi:2010zu}. 
 This new parameter enters the Einstein equations through the energy-momentum tensor, and represents
 the degree of delocalization of the matter distribution 
 \cite{Smailagic:2003yb,Smailagic:2003rp,Smailagic:2004yy,Spallucci:2006zj}.\\
 In this paper we are going to apply the same approach to the axially
 symmetric problem attempting to remove not only the curvature singularity but all the
 pathologies quoted above. 
  
  \section{Preliminary remarks}
It is known that both Schwartzschild and Kerr solutions of general relativity belong to the same class of metrics (\cite{turchi}) with some common properties. Firstly, combinations of the metric components can be brought to the simple form by an appropriate gauge choice \cite{indiano}. Secondly, both metrics can be put in the so called Kerr-Schild form \cite{ks}
i.e. it can be written in terms of the Minkowski metric plus terms involving a specific null vector $k_\mu$. This parametrization has the advantage that the Einstein equations  are linearized in a true sense (not an approximation) which renders them much more tractable. In the course of the paper, we shall exploit these common properties, when needed,  in order to establish analogy between  the two metrics.   This could be particularly helpful having in mind that solving field equations for the Kerr metric is either extremely cumbersome \cite{indiano}, or based on some mathematical procedure  without clear physical input \cite{adler,turchi,Newman:1965tw}.\\ 
Textbook approach introduces Schwarzschild and Kerr geometry as ``vacuum solutions'' of the Einstein equations, where the resulting spacetime symmetry  is an initial assumption. Against this background, our approach follows the basic Einstein's idea that  spacetime is curved due to the presence of  matter.  Consequently, the symmetry of the metric is determined by the symmetry of the matter source. Having at hand the details of the spherically symmetric solution \cite{Nicolini06} we trace the  pattern to follow in this paper. \\
As an introduction of the idea, let us start from the simple Minkowski line element 
written in a spherical basis 

\begin{equation}  
ds^2_M = (dx^0)^2 - dr^2 -r^2 d\vartheta^2 - r^2 \sin^2\vartheta d\phi^2
\end{equation} 

One notices that specific combinations of metric components can be built up to give
 
\begin{eqnarray}
&&\eta_{\vartheta\vartheta}\eta^{rr}=r^2 \label{c1}\\
&&\eta_{00}\eta_{\phi\phi}=-r^2\sin^2\vartheta \label{c2}
\end{eqnarray}

Equations (\ref{c1}), (\ref{c2}) are elementary in Minkowski space, but turn out to be very useful
for black hole spacetime (see (\ref{c5}), (\ref{c6}), because they allow a very simple generalization
leading to a quite non-trivial metric.\\
When matter is present, Schwarzschild-like class of metrics, in the Kerr-Schild form, read

\begin{eqnarray}
&& ds^2_S= ds^2_M -\frac{f(r)}{r^2} (k_\mu dx^\mu)^2\ ,\label{kerr}\\
&& k_\mu=\left(\, 1\ , -1\ , 0\ ,0\,\right)\\
&&k_\mu\,k^\mu=0
\end{eqnarray}

We allow $f(r)$ to be an arbitrary function of the radial coordinate  in order to account for both ordinary (singular), as well as, our regular solution. 
Standard ``vacuum'' solution is 

\begin{equation}
f(r)= \mathrm{const.}\times r
\end{equation}

At this point one cannot but  assign  \textit{a posteriori} significance to the arbitrary
constant by comparing the weak-field with the Newtonian potential. On other other hand, if one starts with a  matter source, i.e. a proper energy
momentum tensor in the r.h.s. of the field equations \cite{Nicolini06,Ansoldi07}, it is \textit{physics} that determines the solution free of any arbitrariness. We found 
 
\begin{equation}
f(r) \equiv 2M(r)r= 2 M\, r \frac{\gamma(3/2\ ; r^2/4l_0^2)}{\Gamma(3/2)}
\end{equation}

where, $M$ is the total mass-energy of the source. \\
Furthermore, on general grounds one can always write Schwartzschild-like solution in terms of the radius dependent mass $M(r)$ defined as

\begin{equation}
M(r)=4\pi \int_0^r dx x^2 \rho(\, x\,)
\end{equation}

where $\rho(r)$ is energy density of matter. The combinations (\ref{c1}), (\ref{c2}) turn out to 
general's in a simple way, as follows

\begin{eqnarray}
&&g_{\vartheta\vartheta}g^{rr}=r^2-f(r)\equiv \Delta(r)\ , \label{c3}\\
&&g_{00}g_{\phi\phi}=-\Delta(r)\sin^2\vartheta \label{c4}
\end{eqnarray}

Now, let us proceed and change the symmetry of the problem. Instead of spherical symmetry  we choose an axially symmetric spheroidal geometry parametrized by the coordinates 

\begin{eqnarray}
&& x=\sqrt{R^2 + a^2}\sin\vartheta \cos\phi\ ,\\
&& y=\sqrt{R^2 + a^2}\sin\vartheta \sin\phi\ ,\\
&& z=R\cos\vartheta \ ,
\end{eqnarray}

Although the above parametrization is found in many  textbooks,
its geometrical meaning is seldom clear, mainly due to the  habit to use a notation which is often indistinguishable from the spherical one. For the reader's advantage, it is worth clearing any possible misinterpretation. Notice that the surfaces described by these coordinates are a family of confocal ellipsoids,for  $R=const.$,  and confocal hyperboloids, for $\vartheta=const$, with foci on the ring $(0\ , a\cos\phi , a\sin\phi\ ,0\,)$.  These surfaces are described by 

\begin{eqnarray} 
&& \frac{{x^2  + y^2 }}{{R^2  + a^2 }} + \frac{{z^2 }}{{R^2 }}= 1\nonumber\\
 &&\frac{{x^2  + y^2 }}{{a^2 \sin ^2 \vartheta }} - \frac{{z^2 }}{{a^2 \cos ^2 \vartheta }} = 1 
 \end{eqnarray}
 
The asymptote of the hyperbola, taking $y=0$ for simplicity, is
 
 \begin{equation}
 z=x\cot\vartheta 
 \end{equation}

Therefore, $\vartheta$ is the angle {\bf between the $z$-axis and the asymptote}. Likewise, coordinate $R$ is the smaller semi-axis of the ellipse, and \textbf{not} the radial coordinate $r$. 
Any function $F(R)$ is \textbf{not} to be considered a radial function in the usual sense.
The parameter $a$ appearing in the Kerr metric  has the geometrical meaning of a focal length.\\
With these preliminary introduction of the symmetry, we write the
Minkowski line element in spheroidal coordinates

\begin{equation}  
ds^2_M = (dx^0)^2 -\frac{\Sigma}{R^2 +a^2} dR^2 -\Sigma d\vartheta^2 - 
(a^2+R^2)\sin^2\vartheta d\phi^2
\end{equation} 

where, $\Sigma\equiv R^2 +a^2\cos^2\vartheta$. Again, we look at specific combinations (\ref{c1}), (\ref{c2}) and find  spheroidal analogue 

\begin{eqnarray}
&&g_{\vartheta\vartheta}g^{RR}=R^2+a^2\equiv \Delta(R)\ , \label{c5}\\
&&g_{00}g_{\phi\phi}=-\Delta(R)\sin^2\vartheta \label{c6}
\end{eqnarray}

The passage from flat to curved space-time is obtained by adding a function $f(R)$ in the definition of $\Delta$.
Thus, we find 

\begin{eqnarray}
&&g_{\vartheta\vartheta}g^{RR}=R^2+a^2-f(R)\equiv \Delta(R)\ , \label{c55}\\
&&g_{00}g_{\phi\phi}=-\Delta(R)\sin^2\vartheta \label{c66}
\end{eqnarray}
 Using (\ref{c55}), (\ref{c66}) one is left  only with
two unknown functions $g_{00}$and $\Delta$, since the component $g_{\vartheta\vartheta}$ is preserved by the spheroidal symmetry. These functions are found using Einstein equations {\textit with appropriate matter source}. However, things can be further simplified exploiting the power of the Kerr-Schild decomposition of the metric. In fact, null four-vector $k_\mu$ can be found solely on the basis of symmetry arguments as

 \begin{eqnarray}
&& ds^2_S= ds^2_M -\frac{f(R)}{\Sigma} (k_\mu dx^\mu)^2\ ,\label{kerr2}\\
&& k_\mu=\left(\, 1\ , -\frac{\Sigma}{a^2+R^2}\ , 0\ ,-a \sin^2\vartheta\,\right)
\end{eqnarray}
 
 One recognizes the above line element as a Kerr-like metric written in the Kerr-Schild form.
 We are left {\textit only} with the unknown scalar function $f(R)$ to be found solving the Einstein equations. It turns out that the Einstein equation for $\Delta$ is particularly simple in form and reads

\begin{equation}
 \frac{d^2\Delta}{dR^2} -2= -16\pi\, g_{\vartheta\vartheta}\left(\, T^R_R + T^\vartheta_\vartheta\,\right)\ ,
 G_N\equiv 1
\label{delta}
\end{equation}

from eq.(\ref{delta}) one deduces, in view of (\ref{c6}), that $ T^R_R + T^\vartheta_\vartheta$ must be of the form

\begin{equation}
 \frac{d^2 f}{dR^2}=16\pi\, g_{\vartheta\vartheta}\, \left(\, T^R_R + T^\vartheta_\vartheta\,\right)
\label{T}
\end{equation}

We hope to have paved a relatively simple way to the generalized Kerr metric without the need to solve complicated equations. Now, we  shall concentrate  on the form of the matter source which produces the above metric and determine the explicit form of the function $f(R)$ through (\ref{T}).

\section{Energy-momentum tensor}

The question of a proper matter source for Kerr (or any other) metric is of paramount importance to give the physical input to Einstein equations. Due to the original ``vacuum approach``,
there have been many attempts to ``engineer''  a suitable form of a matter source to cure
the geometry anomalies. In particular, a general \textit{a posteriori} form of $T^{\mu\nu}$ can be found  by inserting the generalised  axially symmetric,
 stationary, metric into the l.h.s. of the field equations \cite{turchi,Burinskii:2001bq}. In this approach
 energy density and pressures remain unspecified. Various attempts to guess
 suitable matter distributions  reproducing Kerr solution outside  the source and possibly
 regularizing its inner singular behavior  were made 
 \cite{Israel:1970kp,Buri:74,Lopez:1983wm}, \cite{DeLaCruz:1970kk,cohen:1967,Lopez:1984hw},
 \cite{Tiomno:1973ku,Burinskii:2001bq}. 
 However, this has always led to different
 geometries which have to be glued together. Against this background we shall present a 
 unique solution of the field equations, free of any pathology, and smoothly interpolating
 between ordinary Kerr at large distance, and a new regular ``Kerrr'' at short-distance.\\
 In order to determine the energy-momentum tensor we start from the result in the spherically symmetric case \cite{Nicolini06}.

\begin{eqnarray}
&& T^\mu_\nu = \left(\, \rho + p_\theta \,\right)\left(\, u^\mu u_\nu -l^\mu l_\nu\,\right) -p_\vartheta \delta^\mu_\nu\ ,\label{tmn}\\
&& u^\mu =\sqrt{-g_{rr}}\delta^\mu_0\ ,\\
&& l^\mu = -\frac{1}{ \sqrt{-g_{rr}} }\delta^\mu_r\ ,\\
\end{eqnarray}

where, the mass distribution is described by a Gaussian density  $\rho$ as
 
\begin{equation}
\rho(r)=  \frac{M}{(4\pi)^{3/2} l_0^3}\, e^{-r^2/4l_0^2}\label{gauss}
\end{equation}

 The total mass $M$ is defined as the volume integral 

\begin{equation}
M=2\pi \int_0^\infty dr \int_0^\pi d\theta \,\sin\theta \rho(r)
\end{equation}

The above energy momentum tensor is describing an anisotropic fluid and has the same form as in \cite{turchi}.  The pressure $p_\theta $ is determined from the vanishing of the covariant divergence for the energy momentum  tensor \cite{Nicolini06}, which gives

\begin{eqnarray}
&&\partial_r T^r_r=\left(\, T^\theta_\theta -T^r_r\,\right)\, \partial_r \ln g_{\theta\theta}\Rightarrow p_\theta=\rho+\frac{r}{2}\partial_r\rho\label{ptheta}\\
&& \cot{\theta}\left(\, T^\theta_\theta -T^\phi_\phi\,\right)=0\Rightarrow T^\theta_\theta 
=T^\phi_\phi\label{pfi}
\end{eqnarray}

where,  the equation of state $p_r =-\rho$ is understood.
This form of the energy-momentum tensor can  be extended  to the axial symmetry by maintaining its form but changing the explicit expression for pressures and density.  Furthermore, the four velocity $u^\mu$ develops a non-vanishing component $u^\phi$ describing  the rotation of the source. Additional component of $u^\mu$ can be obtained from $u_\mu\,u^\mu=1$ and  gives

\begin{equation}
\frac{u^\phi}{u^0}=\frac{a}{a^2+R^2}\equiv \omega(R)
\label{angv}
\end{equation}

This is the angular velocity of  fluid layers rotating around $z$ axis. 
The mass density $\rho_M(R) $ is now chosen following the reasoning in \cite{turchi,Dymnikova:2006wn}
to pass from non-rotating to rotating physical situation
\begin{equation}
\rho_M\left(\,R\ ,\vartheta\,\right)\equiv   \frac{R^2} {\Sigma}\, \rho_G\left(\,R\,\right)=
\frac{M}{8\pi^{3/2} l_0^3}\, \frac{R^2} {\Sigma}\, e^{-R^2/4l_0^2} 
\label{rhok}
\end{equation}

With the choice (\ref{rhok}) the total mass $M$, defined as the spheroidal volume integral, is
found to be 

\begin{equation}
M\equiv 2\pi \int_0^\infty dR \int_0^\pi d\vartheta \,\sin\vartheta \,\Sigma\, 
\rho_M\left(\,R\ ,\vartheta\,\right)=4\pi \int_0^\infty dR \, R^2\, \rho_G\left(\,R\,\right)
\end{equation}

Finally, the energy-momentum tensor for the generalized Kerr metric is given by

\begin{eqnarray}
&& T^\mu_\nu = \left(\, \rho + p_\vartheta \,\right)\left(\, u^\mu u_\nu -l^\mu l_\nu\,\right) -p_\vartheta \delta^\mu_\nu\ ,\label{tmnk}\\
&& u^\mu =\sqrt{-g^{RR}}\left(\, \delta_\mu^0+\frac{a}{\left(a^2+R^2\right)} \delta_\mu^\phi\,\right) ,\\
&& l^\mu = -\frac{1}{\sqrt{-g_{RR}}}\delta_\mu^R\ ,
\end{eqnarray}

It is important to keep track of different ``$\rho$'' functions present in this case. In particular, $\rho$ in (\ref{tmnk}) is an \textit{invariant energy density}
$\rho= T_{\mu\nu}u^\mu u^\nu$. It is given  in terms of $\rho_G$ as

\begin{equation}
\rho(R\ ,\vartheta)= \frac{R^4}{\Sigma^2} \rho_G(\,R\,)
\end{equation}
in agreement with \cite{Dymnikova:2006wn}.\\
Again, the pressure $p_\vartheta $ and $p_\phi$ are determined from the vanishing covariant 
divergence condition for the energy momentum tensor: 

\begin{eqnarray}
&& T^\vartheta_\vartheta= T^R_R + \frac{\Sigma}{2R}\partial_R T^R_R
\rightarrow -p_\vartheta=\rho+\frac{\Sigma}{2R}\partial_R\rho\\
&& 0= \partial_\vartheta T^\vartheta_\vartheta + 2\cot{\vartheta}\left(\, T^\vartheta_\vartheta -T^\phi_\phi\,\right)\Rightarrow p_\phi=p_\vartheta+\frac{\tan\vartheta}{2}\partial_\vartheta\,p_\vartheta\\
\label{TK}
\end{eqnarray}

which reproduces the corresponding quantities in (\ref{ptheta}), (\ref{pfi})  in the limit
$a\to 0$.

\section{Regular Kerr-like solution}
Our ``Kerrr'' metric reads
 \begin{eqnarray}
ds^2 &&=\left(\, 1 -\frac{2RM(R)}{\Sigma}\,\right)\, dt^2 +\frac{4aM(R)R}{\Sigma}\,\sin^2\vartheta \, dt\, d\phi
 +\nonumber\\
 && -\frac{\Sigma}{\Delta}\, dR^2- \Sigma\, d\vartheta^2 -\frac{\sin^2\vartheta}{\Sigma}\,\left[
 \,\left(\, R^2 + a^2\,\right)^2 -a^2\, \sin^2\vartheta\,\Delta\,\right]\, d\phi^2 \ ,\\
\label{regkerr}
 \Delta &&= R^2 -2M\left(\,R\,\right)R +a^2
 \end{eqnarray}
 
where, $M(R)$ is found to be

\begin{equation}
M\left(\,R\,\right) \equiv 4\pi \int_0^R dx \, x^2\, \rho_G\left(\,x\,\right)
= \frac{M}{\Gamma(3/2)}\, \gamma\left(\, 3/2\ ;R^2/4l_0^2\,\right)
\label{massa}
\end{equation}
 $ \gamma\left(\, 3/2\ ; x\,\right)$ is the lower incomplete gamma function defined as
 
 \begin{equation}
 \gamma\left(\, b\ ; x\,\right)\equiv \int_0^x dt t^{b-1} e^{-t}
\end{equation}
We see that the solution for $M(R)$ has the same form as in the spherically
 symmetric case with the substitution $r\to R$.\\
 
In the above formula $l_0$ is a \textit{minimal length} which, in our approach 
\cite{Smailagic:2003rp,Smailagic:2003rp,Smailagic:2004yy}, is reminiscent of
the underlying non-commutativity of spacetime coordinates leading to the Gaussian matter distribution.\\

On more general ground, $l_0$ can be considered as the width of the Gaussian matter distribution of the source.
Thus, in spite of the  origin of $l_0$, the model is applicable both micro black holes and
astrophysical objects.\\
 
\begin{figure}[h!]
\begin{center}
\includegraphics[width=6cm,angle=0]{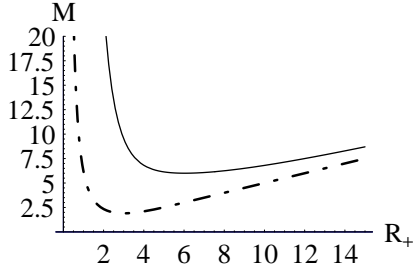}
\caption{\label{kerrr_gr2} 
\textit{ Plot of the function $M(R_H)$ for different values of $\mathbf{a}$ in $l_0=1$ units. The dotted curve corresponds to $\mathbf{a=0}$ (regular Schwarzschild solution), and the continuous curve to $\mathbf{a=6}$. For any $\mathbf{a}$ there is a curve
whose intersections with the line $M=const.$  determines the position of horizons. The minimum corresponds to the extremal black hole. Increasing $\mathbf{a}$ lifts the minimum upwards.}}
\end{center}
\end{figure}

Horizons in (\ref{regkerr}) are real solutions of the equation
\begin{equation}
R^2_H+a^2 -\frac{2M R_H}{\Gamma(\frac{3}{2})}\gamma\left(\,\frac{3}{2} \ ; \frac{R^2_H}{4l_0^2}\,\right) =0 \label{masseq}
\end{equation}
This equation cannot be solved explicitly for $R_H= R_H(\, M\ ; a\,)$ as it is possible for ordinary
Kerr solution.  Thus, we follow an alternative approach: we solve the parameter $M$ in equation
(\ref{massa}) as a function of the horizon radius $R_H$. The plot is given in figure(\ref{kerrr_gr2}).
As in ordinary Kerr solution for any assigned value of $a$ we find three possible situations
\begin{enumerate}
\item $M> M_{extr.}$ there are two distinct horizons $R_\pm$ and the solution represents a non-extremal black hole;
\item$M= M_{extr.}$ there is a single degenerate horizon and the solution correspods to an extremal
black hole of mass 
 given by the minimum value of the curve in figure(\ref{kerrr_gr2}); 
\item $M< M_{extr.}$ there are no horizons. Black hole cannot be formed but there is no curvature singularity, nontheless.
\end{enumerate}

It is important to consider the asymptotic form of our metric (\ref{regkerr})
in $\vartheta=\pi/2$, and $R\to 0$ where ordinary Kerr solution exhibits the infamous
ring singularity.
First, notice that the mass asymptotic behavior  is given by 
\begin{equation}
 M\left(\,R\,\right)\sim \frac{M }{ 6\sqrt{\pi}}\frac{R^3}{l_0^3}\equiv \frac{\Lambda}{6}\, R^3
 \ ,\quad R\to 0
\end{equation}

and

\begin{equation}
\Delta \to  \Delta_\Lambda= R^2+a^2 -\frac{\Lambda}{3}\, R^4
\end{equation}

leading to the rotating deSitter geometry

\begin{equation}
 ds^2 =R^2\,\frac{dR^2}{\Delta_\Lambda}
 -\frac{1}{R^2}\,\left[\,
 a\, dt -\left(\, R^2 +a^2\,\right)\, d\phi \,\right]^2
 +\frac{\Delta_\Lambda}{R^2} 
\,\left[\, dt -a \, d\phi\,\right]^2\ ,
\label{rotds}
\end{equation}

characterized by a scalar curvature

\begin{equation}
\mathcal{R}_{dSr}= 4\Lambda \frac{R^2}{\Sigma}
\end{equation}

The singular ring can be reached by sliding along ellipsoids $R =const.> 0$ 
until arriving on the equatorial plane $\vartheta=\pi/2$ and then letting $R\to 0$. 
In this region we find the Ricci scalar to be
 \begin{equation}
\lim_{R\to 0^+} \mathcal{R}\left(\, R\ ,\vartheta=\pi/2\,\right)= \frac{4M}{\sqrt{\pi} l_0^3}=4\Lambda
 \label{riccids}
\end{equation}
which is the constant curvature scalar of a regular rotating deSitter geometry.
It remains to check what happens when moving along a $\vartheta=const.<\pi/2$ hyperbolae ,
which brings us down on equatorial disk with $R=0$. In this case, we find 

  \begin{equation}
\lim_{R\to 0^+} \mathcal{R}\left(\, R\ ,\vartheta_0<\pi/2\,\right)= 0
\label{riccimink}
\end{equation}
and
\begin{equation}
\lim_{R\to 0^+} \rho\left(\, R\ ,\vartheta_0<\pi/2\,\right)= 0
\end{equation}
Thus, the disk is a matter-free, zero curvature   Minkowski flat spacetime, as it is in the ordinary Kerr geometry:

\begin{equation}  
ds^2_M = -(dt)^2+ \cos^2\vartheta_0 \, dR^2 + a^2\sin^2\vartheta_0 d\phi^2
\label{m2}
\end{equation} 

The difference with ordinary Kerr  is that the  singular ring is replaced by
a regular deSitter, Saturn-like region of non-zero width, with inner radius $x^2+y^2=a^2$. 
Our model represents the first explicit  example of a matter source leading to a singularity-free
metric that naturally interpolates between near-by de Sitter and outer Kerr-like forms. No ad hoc conjectures, or patching, is required. 

\section{The stringy heart of the Kerrr solution}

From (\ref{riccids}) and (\ref{riccimink}) we see that there is a discontinuity
in the Ricci scalar as one approaches $R\to 0\ , \vartheta\to \pi/2$ from two different directions.
One may wonder wether thia jump has a physical meaning?
We shall try to answer this question.\\
First notice that the metric induced on the equatorial  plane is strongly reminiscent of the spacetime
geometry in the presence of a vacuum bubble \cite{Berezin:1987bc}. 
To be more precise, we can intepret the Minkowski
disk as a ``true vacuum'' \textit{planar bubble} surrounded by a deStter ``false vacuum'' and we can apply the Israel matching condition \cite{Israel:1966rt} to give a physical meaning to the metric discontinuity. We write the flat
metric (\ref{m2}) in terms of planar polar coordinates  $r=a\sin\vartheta$, $\phi$, as

\begin{equation}
 ds^2_{in}= dt^2 -dr^2 - r^2d\phi^2\ , r\le a
\end{equation}

In the same way we write the outer equatorial plane,outer, near-ring deSitter geometry as

\begin{equation}
ds^2_{out}= \left[\, 1 - \frac{\Lambda}{3}\left(\, r^2-a^2\,\right)\,\right]\,   dt^2 
-dr^2 - r^2d\phi^2\ , r\ge a
\end{equation}

The matching of the extrinsic curvatures along the static boundary $r=a$

\begin{equation}
 \epsilon_{in} \sqrt{\eta_{00}} -\epsilon_{out} \sqrt{g_{00}(r=a)} = 4\pi G_N\, \sigma
\end{equation}

where, $\epsilon_{in/out}=\pm 1$ according with the choice of orientation of the normal to the ring; 
 $\sigma>0$ is the linear energy density, i.e. the \textit{tension}, of the ring. The jump in
 the extrinsic curvature is non-zero for $\epsilon_{in}=1$, $\epsilon_{out}=-1$ leading to
 
 \begin{equation}
  \sigma=\frac{1}{2\pi G_N}\equiv \frac{1}{2\pi \alpha^\prime}
 \end{equation}

 From this expression for the tension we can recover the mass of the ring $M_r$ as
 
 \begin{equation}
 M_r\equiv 2\pi a \sigma = \frac{a}{\alpha^\prime}
 \label{mring}
\end{equation}

Now, we can compute the ring angular momentum which rotates with angular velocity
\begin{equation}
 \omega_r = \omega(R=0)=\frac{1}{a}
\end{equation}

thus, we get

\begin{equation}
 J_r = M_r a^2 \omega_r = M_r a \label{jring}
\end{equation}

By inserting (\ref{jring}) in (\ref{mring}) we find

\begin{equation}
 J_r= \alpha^\prime M_r^2
 \label{regge}
\end{equation}

which is a classical {Regge Trajectory} with a planckian Regge slope $\alpha^\prime = M_{Pl.}/l_{Pl.}$  !\\
This result offers an exciting interpretation of the ring as a \textit{classical, rotating, circular string}
leaving on a Regge trajectory.  Therefore, we offer the folowing physical interpretation of our solution.
The ``heart'' of the  Kerrr black hole is a rotating string of finite  tension replacing the standard
Kerr singularity. The stringy interpretation is supported by the Regge realtion between the mass and
the angular momentum of the ring.
The string is immerged in a cloud of matter described by the $T^{\mu\nu}$ discussed
previously and the is inner engine inducing rotation of the matter elipsoidal layers.
The gaussian profile of the matter cloud is instrumental to regularize the inifinte curvature jump, present in the ordinary Kerr solution, to a finite value $4\Lambda$, thanks to the presence of the outer deSitter belt. 
Or, in other words, the infinte tension ring-like curvature singularity is ``renormalized`` to
the the maximum physically acceptable Planckia tension of a fundamental string. 
 
\section{Diving through the equatorial disk}
There is another question long awaiting a satisfactory answer in the Kerr metric.
It is known that, unless it is forcefully cut-off, there is an ``anti-gravity'', negative $R$ region, were causality violation takes place due to the existence  of closed time-like curves. We shall show that our metric resolves both problems. Let us go to the $z$-axis by taking $\vartheta=0$ and $R=|z|$. Then,  $g_{00}$ reads

\begin{equation}
 g_{00}= 1 - \frac{2 |z| M(z)}{z^2+a^2} 
\end{equation}
\begin{figure}[h!]
\begin{center}
\includegraphics[width=6cm,angle=0]{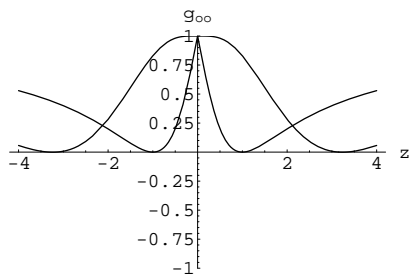}
\caption{\label{kerrr_gr3} 
\textit{ Plot of $g_{00}$ along the axis of rotation for Kerr (spiky curve) 
and Kerrr (smooth curve) solutions.}}
\end{center}
\end{figure}
The problem of multiple Riemannian sheets arises whenever $g_{00}$ is a function  of odd power of $|z|$. In the usual Kerr $M(z)=const.$ and $g_{00}$  is discontinuous in the first derivative at $z=0$ leading to a jump
in the extrinsic curvature. In our case, the behavior of $M(z)$ near the origin is given by

\begin{equation}
M(z)\sim const.\times z^3 
\end{equation} 

which gives an {\textit {even}} function $g_{00}$ for small argument with a continuous first derivative
in $z=0$. In the standard Kerr geometry ther are two different interpretations of the discontinuity
of the gravitational field across the Minkowski disk:
\begin{enumerate}
 \item discontinuity is justified by the presence of matter on the disk. Unfortunately, this matter  turns 
 out to move in a  \textit{super-luminal} fashion and  have divergent density on the ring  \cite{Israel:1970kp,hamity}.
\item Discontinuity is removed by interpreting the disk as a ``branch cut'' and attaching a second
Riemann sheet of negative $R$. This procedure restores analiticity at the price of introducing a negative gravity sheet of the metric, where closed time-like curves can exist as a consequence 
of allowing $R$ to be negative. 
\end{enumerate}

Our solution has no problems of this kind since it is analytic everywhere, and thus it is meaningless to talk about analytic continuation of the metric. In other words, geodesics can cross
the Minkowski disk without any problem.
Everything fits nicely together, as conjectured, following the same line of reasoning already encountered  in the regular spherically symmetric case.

\section{Conclusions}
In this paper we presented the first example of a smooth  matter distribution which
leads to a pathology-free Kerr solution. The form of the source  is a generalization
of the corresponding Gaussian mass/energy distribution we introduced for spherically
symmetric sources  to the case of a rotating object. For both solutions, the same mechanism is at work: the curvature singularity is replaced by a deSitter vacuum domain. In the spherically symmetric case it is an inner deSitter core, while
in the Kerrr solution it turns out to be a Saturn-like  belt of rotating deSitter vacuum, surrounding an empty Minkowskian disk. The novel feature of the Kerrr solution is  that the Minkowski disk
joins the deSitter belt through a  a rotating string with Planckian (finite!) tension. \\
Beside removing the nasty ring singularity the gaussian cloud of matter eliminates the negative $R$
sheet of the Kerr black hole by ensuring analyticity of the metric across the disk. Positivity
of $R$ forbids the presence of closed time-like curves.\\
To keep the length of the paper short enough to fit the journal format,
we must postpone a detailed study  of  Kerrr black hole thermodynamics to a next article.
We anticipate that as in the case of spherically symmetric regular black holes, we find that the Hawking
temperature is not unbounded but reaches a maximum value and then drops to zero as the extremal
configuration is approached.\\
In a forthcoming paper we shall present the extension of the present work to the Kerr-Newman black hole.

\end{document}